\documentclass[12pt]{iopart}
\usepackage{epsfig}

\newcommand{\beq}{\begin{eqnarray}}
\newcommand{\eeq}{\end{eqnarray}}
\newcommand{\be}{\begin{eqnarray*}}
\newcommand{\ee}{\end{eqnarray*}}

\def\lsim{\raise0.3ex\hbox{$<$\kern-0.75em\raise-1.1ex\hbox{$\sim$}}}
\def\gsim{\raise0.3ex\hbox{$>$\kern-0.75em\raise-1.1ex\hbox{$\sim$}}}

\begin{document}

\title[Heavy-quark production from Glauber-Gribov theory at
LHC]{Heavy-quark production from Glauber-Gribov theory at LHC}

\author{I.C. Arsene$^{1}$, L. Bravina$^{1,2}$, A.B. Kaidalov$^{3}$, K.
  Tywoniuk$^{1}$ and E. Zabrodin$^{1,2}$}

\address{$^1$ Department of Physics, University of Oslo, N-0316 Oslo}
\address{$^2$ Institute of Nuclear Physics, Moscow State University,
  RU-119899 Moscow}
\address{$^3$ Institute of Theoretical and Experimental Physics,
  RU-117259 Moscow}
\ead{konrad@fys.uio.no}
 
\begin{abstract}
We present predictions for heavy-quark production for proton-lead
collisions at LHC energy 5.5 TeV from Glauber-Gribov theory of
nuclear shadowing. We
have also made predictions for baseline cold-matter (in other words
inital-state) nuclear effects in
lead-lead collisions at the same energy that has to be taken into
account to understand properly final-state effects.
\end{abstract}

\maketitle

\section{Introduction}
In the Glauber-Gribov theory \cite{Gri69} nuclear shadowing at low-$x$ is
related to diffractive structure functions of the nucleon, which are
studied experimentally at HERA. The space-time picture of the
interaction for production of a heavy-quark state on nuclei changes
from longitudinally ordered rescatterings at energies below the
critical energy, corresponding to $x_2$ of an active parton from a
nucleus becoming smaller than $1/m_N R_A$, to the coherent interaction
of constituents of the projectile with a target nucleus at energies
higher thant the critical one \cite{Bor93}. For production of $J/\psi$ and
$\Upsilon$ in the central rapidity region the transition happens at
RHIC energies. In this kinematical region the contribution of
Glauber-type diagrams is small and it is necessary to calculate
diagrams with interactions between pomerons, which, in our approach,
are accomodated in the gluon shadowing. A similar model for
$J/\psi$-suppression in d+Au collisions at RHIC has been considered in
Ref.~\cite{Cap06}.

Calculation of gluon shadowing was performed in our recent paper
\cite{Ars07}, where Gribov approach for the calculation of nuclear
structure functions was used. The gluon diffractive distributions were
taken from the most recent experimental parameterizations of HERA data
\cite{H106}. The Schwimmer model was used to account for higher-order
rescatterings. 

\section{Heavy-quark production at the LHC}
\begin{figure}[t!]
  \begin{minipage}[t]{1.\linewidth}
    \begin{center}
      \includegraphics[scale=.5]{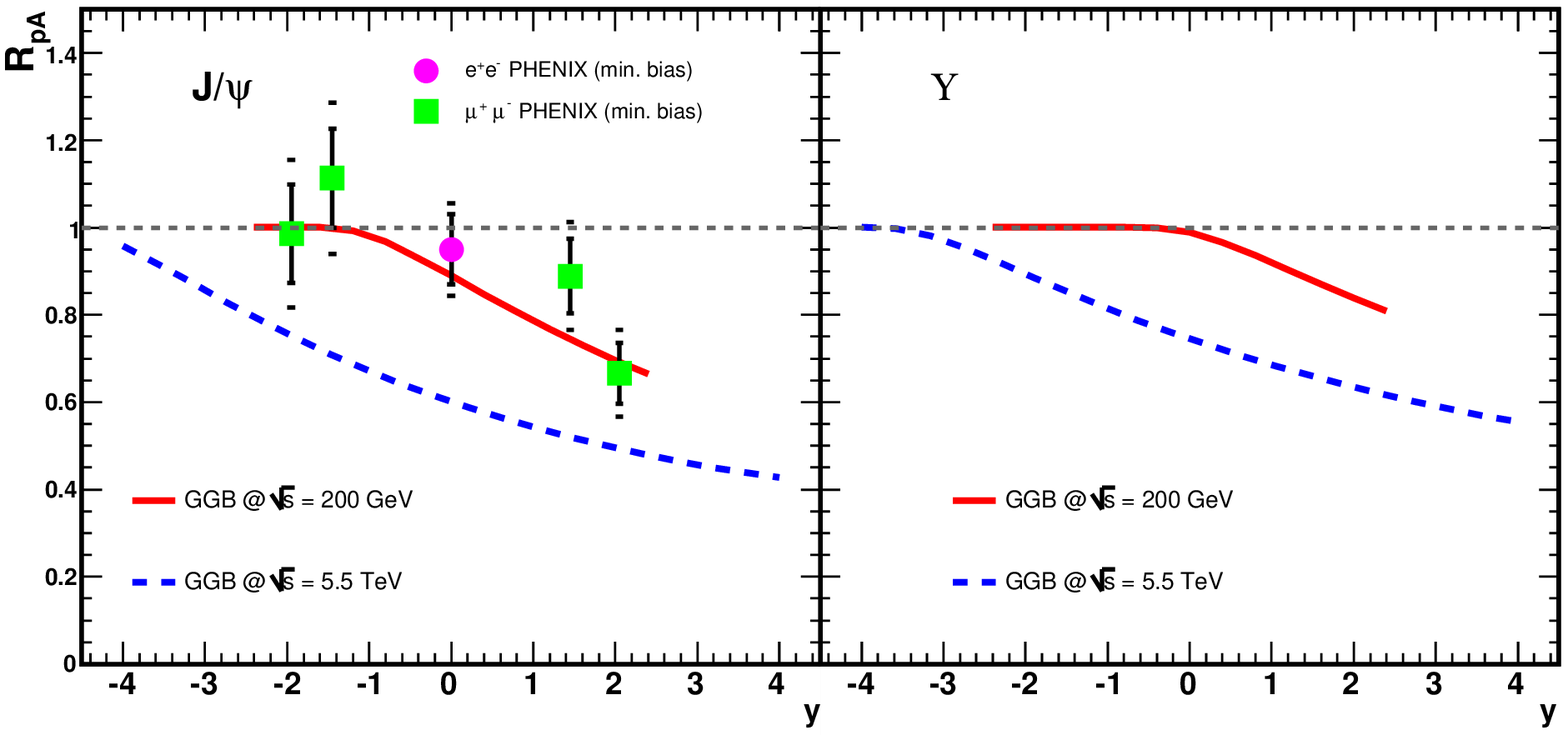}\\
      \includegraphics[scale=.5]{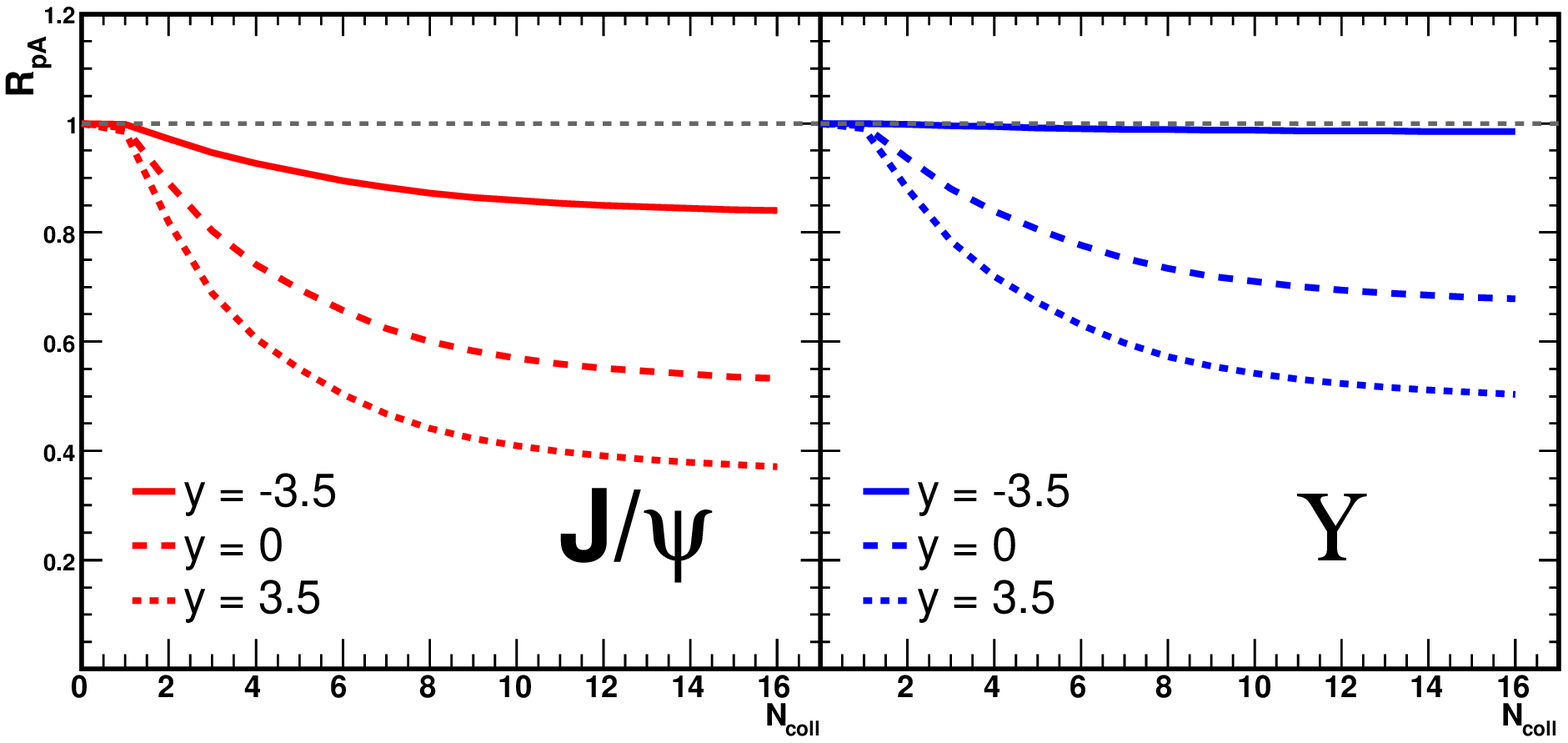}
    \end{center}
  \end{minipage}
  \caption{Rapidity (top) and centrality (bottom) dependence of the
    nuclear modification factor for 
    $J/\psi$ (left) and $\Upsilon$ (right) production in p+Pb (d+Au)
    collisions at $\sqrt{s} =$ 5500 (200) GeV. Experimental data are
    from \cite{PHE03}.}
  \label{fig:RapDep}
\end{figure}

We present predictions for the rapidity and centrality dependence of
the nuclear modification factor in proton-lead (p+Pb) collisions for
both $J/\psi$ and $\Upsilon$ in 
Fig.~\ref{fig:RapDep} (the data on $J/\psi$ suppression at $\sqrt{s} =
200$ GeV is taken from \cite{PHE03}, where also a 
definition of the nuclear modification factor can be found). We
predict a similar suppression for open charm, $c\bar{c}$, and bottom,
$b\bar{b}$, as for the hidden-flavour particles.
The observed $x_F$ scaling at low energies of the parameter $\alpha$
(from $\sigma_{pA} = \sigma_{pp} A^\alpha$) for $J/\psi$ production,
which is broken already at RHIC, will go to a scaling in $x_2$ at
higher energies. This will also be the case for $\Upsilon$ and open
charm and bottom. 

\begin{figure}[t]
  \begin{center}
    \includegraphics[scale=.4]{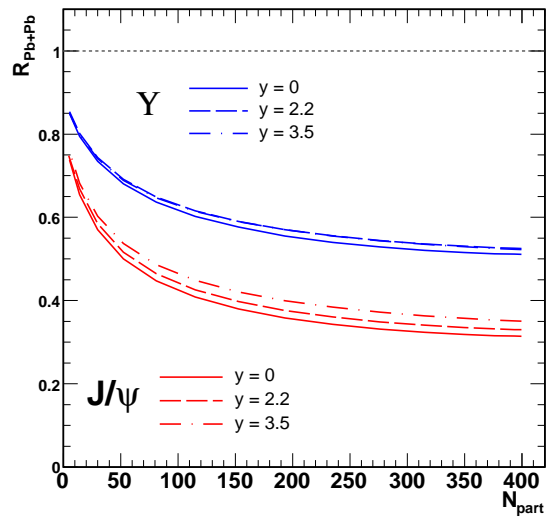}
  \end{center}
  \caption{Baseline cold-nuclear matter effects in Pb+Pb collisions at
    5.5 TeV for $J/\psi$ and $\Upsilon$ production.}
  \label{fig:ColdNucl}
\end{figure}

In Fig.~\ref{fig:ColdNucl} we present predictions for cold-nuclear
matter effects due to gluon shadowing in lead-lead (Pb+Pb) collisions
at LHC energy $\sqrt{s} = 5.5$ TeV for the production of $J/\psi$ and
$\Upsilon$. The suppression is given as a function of rapidity and
centrality. . 

\end{document}